\documentclass[aps,pra,showpacs,superscriptaddress,superscriptaddress,twocolumn]{revtex4-1}  
\usepackage{graphicx}  
\usepackage{dcolumn}   
\usepackage{bm}        
\usepackage{amssymb}   
\usepackage{amsmath}
\usepackage{subfigure}
\usepackage{epstopdf}
\begin{document}


\title{Spatial correlations of one dimensional driven-dissipative systems of Rydberg atoms}

\author{Anzi Hu}
\affiliation{Joint Quantum Institute, University of Maryland and National Institute of Standards and Technology}
\author{Tony E. Lee}
\affiliation{ITAMP, Harvard-Smithsonian Center for Astrophysics, Cambridge, MA 02138, USA}
\author{Charles W. Clark}
\affiliation{Joint Quantum Institute, University of Maryland and National Institute of Standards and Technology}

\date{\today}
\begin{abstract}
We consider a one-dimensional lattice of atoms with laser excitation to a Rydberg state and spontaneous emission. The atoms are coupled due to the dipole-dipole interaction of the Rydberg states. This driven-dissipative system has a broad range of non-equilibrium phases, such as antiferromagnetic ordering and bistability. Using the quantum trajectory method, we calculate the spatial correlation function throughout the parameter space for up to 20 lattice sites. We show that bistability significantly strengthens the spatial correlations and the entanglement.
\end{abstract}
\maketitle

\section{Introduction}

A current challenge is to understand quantum many-body systems in the presence of driving and dissipation. Such systems converge to a steady state, given by the balance of coherent and dissipative evolution. Classical driven-dissipative systems are known to display a broad range of intriguing spatiotemporal order \cite{PerBak1988, Cross1993}. The question is then what happens in quantum driven-dissipative systems. Although one might expect dissipation to lead to trivial states, it has been shown that ordered phases can exist for systems with quasi-local dissipative mechanisms \cite{diehl2008quantum,Prosen2011, YiDaleyZoller_2012, hoening_Fleischauer2012}, in the presence of 1/f noise \cite{DemlerGiamarchiAltman2010quantum}, and for a quadratic Hamiltonian with local dissipation \cite{Prosen2010noise}. These works show that it is possible for nontrivial states to exist even when dissipation is present, although they either assume a dissipation mechanism that creates coherence between neighbors or a noise process of a particular form.

A convenient setting to study driven-dissipative many-body phenomena is Rydberg atoms \cite{ates2007,pritchard2010,weimer10,Lee_2012_AntiF,Lee2012_Jump,Lesanovsky2012,qian2012,Lee2012_threeRyd,Pupillo2012,dudin12,peyronel2012,Honing_Fleischhauer2013,carr2013cooperative,malossi13,gorshkov2013,carr_saffman2013,Lesanovsky2013,lemeshko13,lee13,qian13,petrosyan13a,petrosyan13b}. Rydberg states are highly excited electronic states of an atom, and there is a strong dipole-dipole interaction between neighboring Rydberg atoms \cite{lukin2001,saffman10}. The spontaneous decay from the Rydberg state is a natural and inevitable source of dissipation \cite{gallagher94}. Thus, one creates a driven-dissipative system by continuously laser-exciting atoms to the Rydberg state and letting them decay back to the ground state. Mean-field theory for a lattice of Rydberg atoms predicts different behavior for different parameter regions \cite{Lee_2012_AntiF}. For example, there can be antiferromagnetic or crystalline ordering, in which every other atom has higher Rydberg population. Another feature is bistability, in which there are two stable collective states: one with low Rydberg population and one with high Rydberg population. 

An open question is what happens without the mean-field approximation, especially in low dimensions. Simulations of Rydberg atoms in a one dimensional (1D) lattice in the antiferromagnetic region showed that the correlation decays very quickly as a function of distance, implying the absence of long-range order \cite{Lee_2012_AntiF,Honing_Fleischhauer2013}. Simulations of a 2D lattice in the antiferromagnetic region found strong classical correlations of Rydberg excitation probabilities \cite{petrosyan13b}. Simulations of the all-to-all coupling model in the bistable region found strong temporal correlations between the atoms: the system collectively jumps between the two stable states \cite{Lee2012_Jump}. Simulations of the 1D model found strong temporal correlations in the bistable region, but the simulation was for a large coupling ($V=100\gamma$) and a system of $N=12$ lattice sites\cite{Lesanovsky2012}. Recently, two experiments have observed bistability in a gas of Rydberg atoms \cite{carr2013cooperative,malossi13}, although the atoms are not fixed on a lattice, and they are not perfect two-level systems.

In this paper, we examine the spatial correlations in 1D more thoroughly by exploring more fully the parameter space and by using larger systems. We find that the correlation decays slowly as a function of distance in the bistable region, much more slowly than in the antiferromagnetic region. This indicates that bistability can significantly enhance correlations in driven-dissipative systems. We also find that bistability increases the amount of entanglement.

\section{Model}

We consider a 1D chain of $N$ atoms continuously excited by a laser from the ground state to a Rydberg state. The atoms are approximated as two-level systems. Let $|g\rangle_j$ and $|e\rangle_j$ denote the ground and Rydberg states of the atom at site $j$. The dipole-dipole interaction of the Rydberg states leads to a level shift $V$ when two atoms are in the Rydberg state. We assume nearest-neighbor interactions, which is a good approximation in the van der Waals regime, when the interaction decays as the sixth power of distance. The dipole-dipole interactions involving ground states are much weaker, so we ignore them. The Hamiltonian in the interaction picture and rotating-wave approximation is written as
\begin{align}
H=&\sum_j \bigg[-\Delta |e\rangle\langle e|_j+\frac{\Omega}{2}(|e\rangle\langle g|_j+|g\rangle\langle e|_j) \nonumber \\
&+V|e\rangle\langle e|_j\otimes|e\rangle\langle e|_{j+1}\bigg],
\label{eq:H}
\end{align}
where $\Delta$ is the detuning of the laser from the Rydberg transition and $\Omega$ is the Rabi frequency. The Hamiltonian can be mapped to a spin-1/2 Ising model with both transverse and longitudinal fields \cite{Lesanovsky2012}. In the corresponding Ising model, the coupling $V$ corresponds to spin-spin coupling along $z$-axis, the drive $\Omega$ corresponds to a transverse magnetic field along the $x$-axis, and $\Delta-V$ corresponds to a longitudinal field along the $z$-axis.

We account for the spontaneous emission from the Rydberg state using the linewidth $\gamma$. We assume that each atom emits into different electromagnetic modes, i.e., the decay is not superradiant. (This approximation is appropriate for relatively small $N$, but for large $N$, superradiant decay to neighboring Rydberg states dominates \cite{carr2013cooperative,wang07,day08}). 

The time evolution of the system is described by the master equation of its density matrix $\rho$,
\begin{equation}
\dot{\rho}=-i[H,\rho]+\gamma\sum_j \left(-\frac{1}{2}\{|e\rangle\langle e|_j,\rho\}+|e\rangle\langle g|_j\, \rho \,|g\rangle\langle e|_j\right).
\label{eq:master}
\end{equation}
In this paper, we follow the convention that all parameters are in the units of the decay rate $\gamma$, and we set $\gamma= 1$. The steady-state density matrix describes the statistical properties of the system. We calculate the correlation between two atoms at sites $i$ and $j$:
\begin{equation}
c_{ij}=\langle E_i E_j\rangle-\langle E_i\rangle\langle E_j\rangle, 
\label{eq:g22}
\end{equation}
where $E_j=|e\rangle\langle e|_j$ is the projection operator for the Rydberg state of atom $j$.



\begin{figure}[t]
\includegraphics[width=0.54\textwidth]{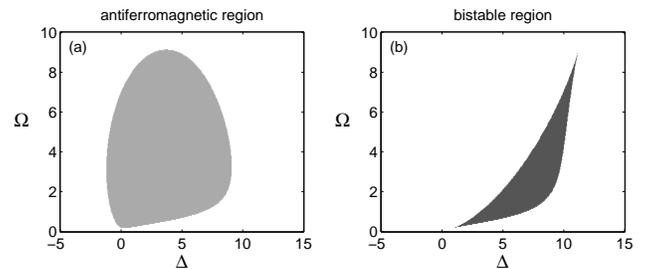}
\caption{Mean field phase diagram as a function of $\Delta$ and $\Omega$ with V=10. The shaded area indicates the antiferromagnetic region in (a) and  the bistable region in (b).}
\label{fig:phase_dia}
\end{figure}

The master equation can be solved using a mean-field approximation \cite{Lee2012_Jump,Lee_2012_AntiF}. There are two parameter regions of interest. When $\Delta\approx 0$, the lattice exhibits antiferromagnetic ordering, where one sublattice of atoms has higher Rydberg population than the other sublattice. This is because the laser is originally on resonance with the Rydberg transition, but the more-excited sublattice shifts the other sublattice off resonance due to the blockade effect \cite{lukin2001}. When $\Delta\sim V$, there is bistability, meaning that there are two steady states: one with low Rydberg population and one with high Rydberg population \cite{Lee2012_Jump,Lesanovsky2012}. The intuition for bistability is as follows. Since the laser is originally off resonance, the atoms will not be very excited. But if the atoms happen to be very excited, the Rydberg interaction effectively shifts the atoms onto resonance with the laser and the atoms remain very excited. Figure \ref{fig:phase_dia} shows the mean-field phase diagram as a function of $\Omega$ and $\Delta$, showing the antiferromagnetic and bistable regions. The phase diagram as a function of $\Omega$ and $V$ is given in \cite{Lesanovsky2012}.

The mean-field approximation does not take into account inter-atomic fluctuations. When one simulates the dynamics without the mean-field approximation, in the bistable region, one finds that the system occasionally jumps between the two mean-field steady states \cite{Lee2012_Jump,Lesanovsky2012}. As a result, the system exhibits bright periods (high rate of photon emission) and dark periods (low rate of photon emission).


\begin{figure}
\includegraphics[width=0.40\textwidth]{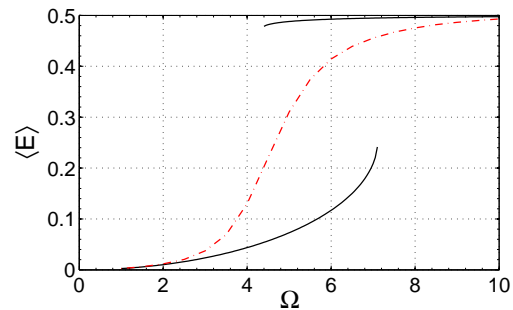}
 \caption{Rydberg population $\langle E \rangle$ as a function of $\Omega$ for $\Delta=V=10$. The black lines denote the stable solutions under the mean-field approximation. The red dashed line is the quantum trajectory calculation for $N=15$. }
 \label{fig:E}
 \end{figure}

 \begin{figure}
 \subfigure[]{\includegraphics[width=0.48\textwidth]{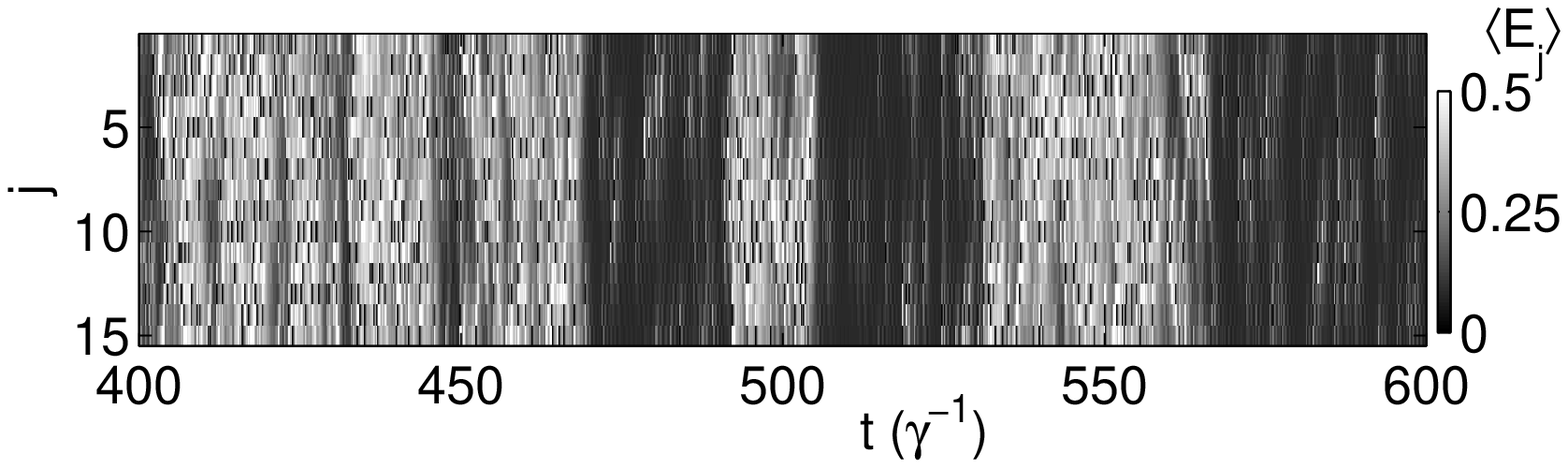} }
 \subfigure[]{\includegraphics[width=0.48\textwidth]{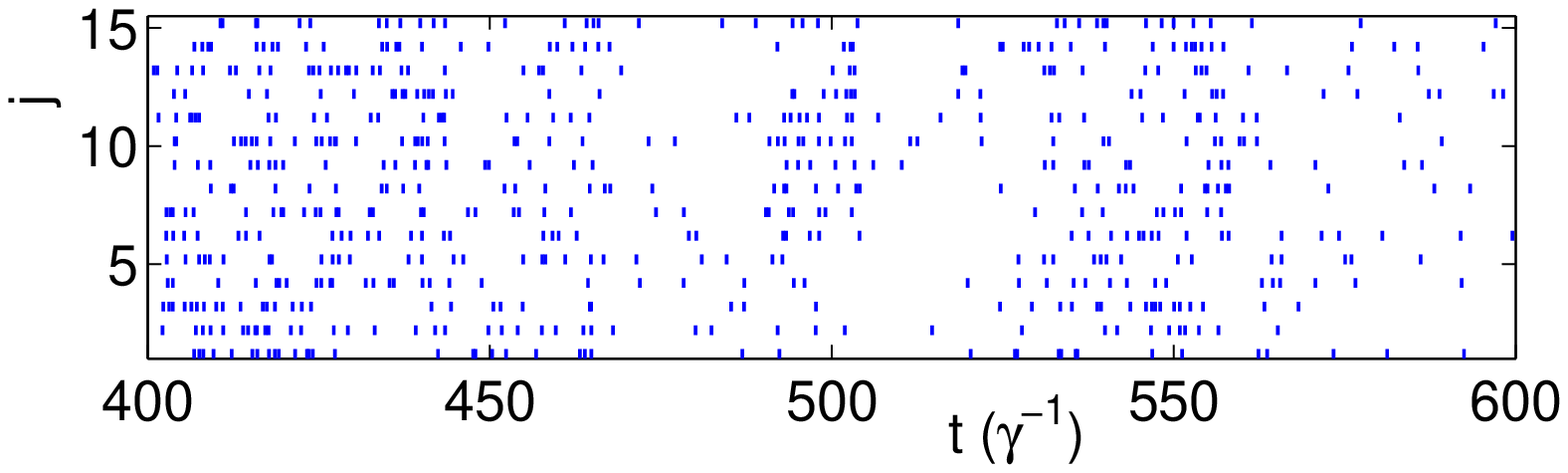}}
\caption{An example trajectory in the bistable region. The parameters are $N=15$, $V=\Delta=10$ and $\Omega=4.5$. (a) $\langle E_j\rangle$ for each atom $j$ as a function of time, using color scale on right. 
(b) Each vertical line corresponds to a photon emission from site $j$ at a given time. }
\label{fig:traj}
\end{figure}

\section{Results}
\subsection{Spatial Correlation}
\begin{figure}[t]
\includegraphics[width=0.43\textwidth]{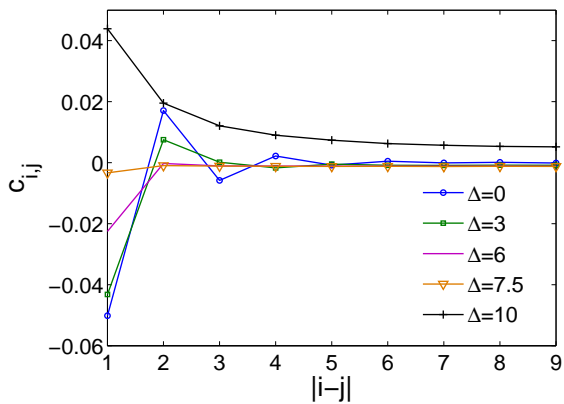}
\caption{Correlation function $c_{i,j}$ for $N=18$, $V=10$, $\Omega=4.5$, and different values of $\Delta$. 
}
\label{fig:g2}
\end{figure}

\begin{figure}[t]
\subfigure[]{\includegraphics[width=0.43\textwidth]{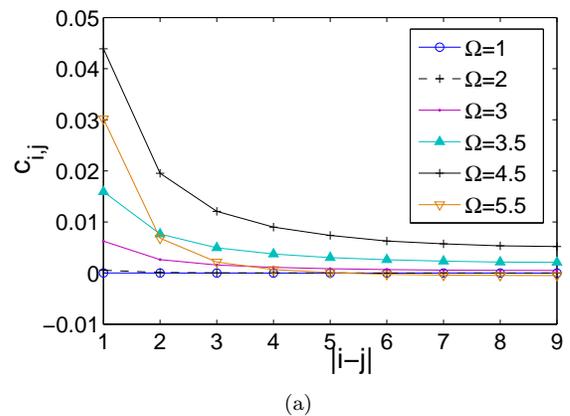}}
\subfigure[]{\includegraphics[width=0.42\textwidth]{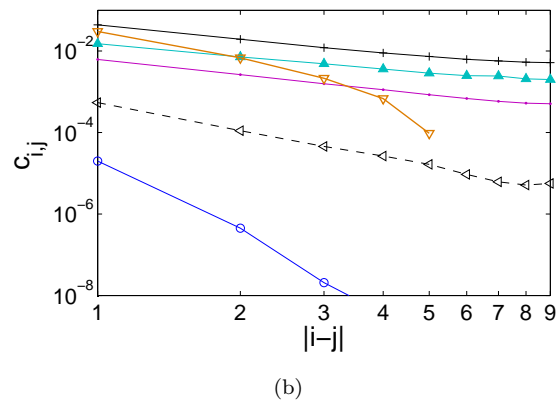}}
\caption{Correlation function $c_{i,j}$ for $N=18$, $V=10$, $\Delta=10$, and different values of $\Omega$ in linear (a) and log-log (b) plots. The correlation for $\Omega=5.5$ has a negative value on the order of $10^{-4}$ for $|i-j|\geq 6$. 
}
\label{fig:g2_log}
\end{figure}

\begin{figure}[t]
\subfigure[]{\includegraphics[width=0.4\textwidth]{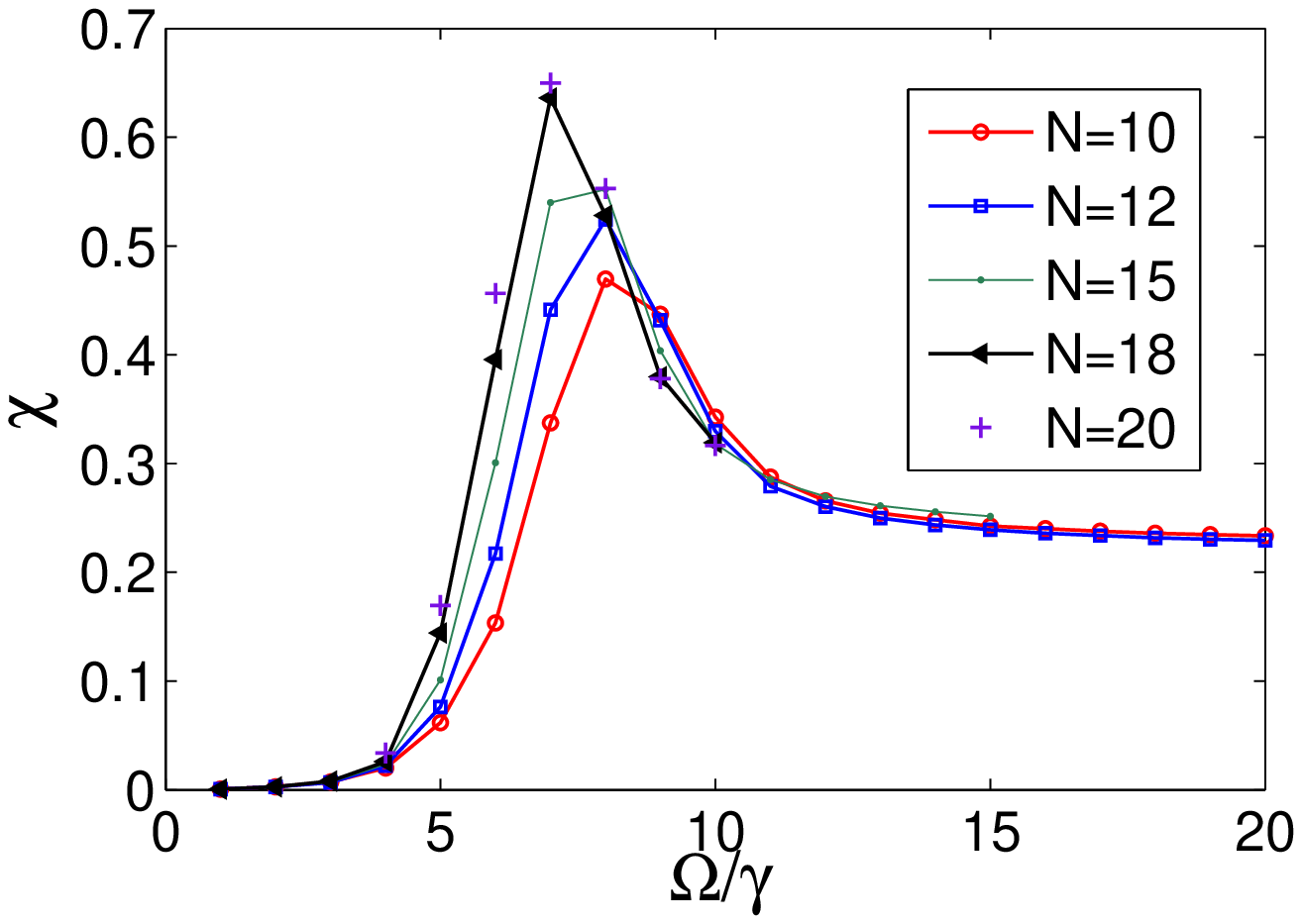}}
\subfigure[]{\includegraphics[width=0.4\textwidth]{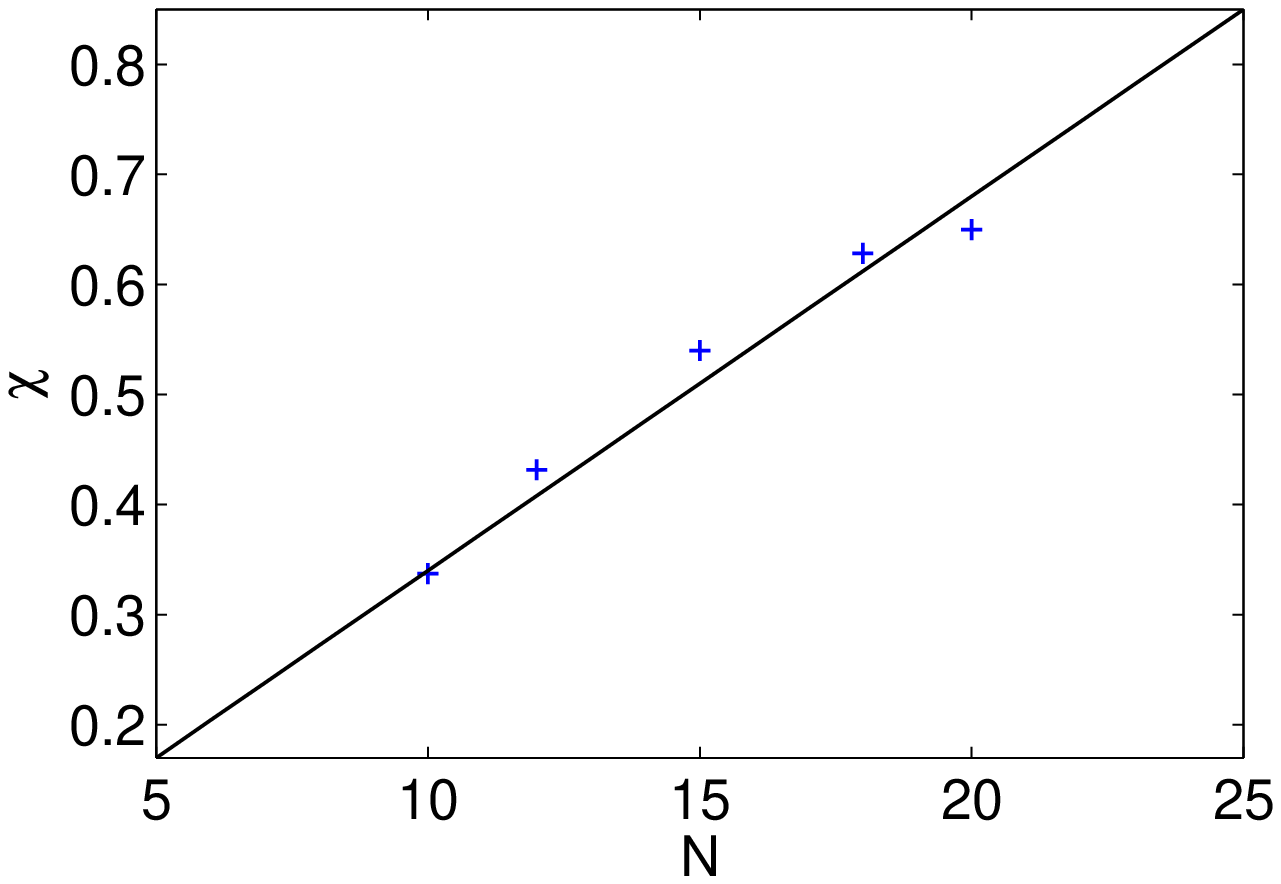}}
\caption{(a) Excitation susceptibility as a function of $\Omega$ for different system sizes for (a) $V=\Delta=20$. (b) Excitation susceptibility as a function of system sizes for $V=\Delta=20$ and $\Omega=7$. The solid line is the fit function $y=0.034x$. 
}
\label{fig:sus}
\end{figure}

We study the quantum dynamics in a 1D lattice using the quantum trajectory method \cite{molmerDalibard1993}, which generates an ensemble of trajectories whose statistical properties reflects the time evolution described by the master equation. Each trajectory in the ensemble can be viewed as a single experimental realization. In our work, time integration between photon emissions is done using the fourth order Runge-Kutta method with a time step of $0.001\gamma^{-1}$. To ensure accuracy, for each parameter setting, we average over $10^{6}$ sample points, at time intervals of $0.01\gamma^{-1}$. With the implementation of parallel computing, we are able to calculate the correlation function $g^{(2)}$ for up to $N=20$. We assume periodic boundary conditions.

We first compare the average Rydberg population $\langle E\rangle$, where $E=N^{-1}\sum_j E_j$, for mean-field and quantum trajectories. Figure \ref{fig:E} plots 
$\langle E\rangle$ as a function of $\Omega$ for fixed $V$ and $\Delta$.
Both calculations gives similar results when $\Omega$ is very small or very large and the system is far from the bistable region. As the system approaches the bistable region, the two calculations begin to deviate. In the region where there are two stable solutions, mean-field calculation gives two different values of $\langle E\rangle$, and the system converges to one of them depending on the initial conditions. However, fluctuations beyond mean-field cause the system to switch between the two states, and the discontinuity is replaced by a smooth increase of $\langle E\rangle$ as $\Omega$ increases. The deviation between the mean-field and quantum trajectory calculation illustrates the importance of fluctuations beyond mean-field. Fig.~\ref{fig:traj} shows an example trajectory. One can clearly see the collective jumps between bright and dark periods as well as cascades of photon emissions from neighboring atoms.

Figs. \ref{fig:g2} and \ref{fig:g2_log} plot the spatial correlation function $c_{i,j}$ as defined in Eq.~\eqref{eq:g22}. We first compare the correlation function in the antiferromagnetic region with the correlation function in the bistable region (Fig.~\ref{fig:g2}). For fixed $V$ and $\Omega$, the system changes from antiferromagnetic to bistable as $\Delta$ increases from zero to $V$ \cite{Lee_2012_AntiF}. In the antiferromagnetic region ($\Delta=0,3$), the correlation shows the alternating sign, but decays rapidly as a function of distance. In the bistable region ($\Delta=10$), the correlation is positive and decays slowly. In fact, the correlation in the bistable region decays much more slowly than in the antiferromagnetic region. Thus, bistability significantly enhances the correlations. 

Fig.~\ref{fig:g2_log} compares the spatial correlation for different values of $\Omega$ when $V$ and $\Delta$ are fixed at 10. The correlation is very small in the ``dark'' state ($\Omega=1,2$). As the drive increases, the correlation grows. The correlation is strongest at around $\Omega=4.5$. Interestingly, this is also where bistability predicted by mean-field theory starts. As the drive increases further ($\Omega=5.5$), the correlation becomes short-range again. To distinguish the slow decaying correlation, we show the spatial correlation in a log-log plot in Fig.~\ref{fig:g2_log}(b). We notice that the correlation changes from the short-ranged exponential decay in the ``dark'' state ($\Omega=1,2$) to slower decay around $\Omega=2.5$ and becomes slower than a power law decay as the system enters the bistable region($\Omega=3,4.5$). The correlation becomes exponential again as the drive increases further ($\Omega=5.5$). The fact that the decay is slower than power law for $\Omega=3,4.5$ is probably a finite-size effect, but it is clear that the correlations in the bistable region are significantly stronger than elsewhere.

The decay of the correlations is due to the presence of domain walls in the system, and the fact that the system is 1D makes it particularly susceptible to domain walls. The fact that the correlation decays slower in the bistable region than the antiferromagnetic region can be intuitively understood in terms of domain walls. Suppose the system is in the antiferromagnetic region and the system starts in a perfect antiferromagnetic state, i.e., every other atom is excited. It is quite easy for domain walls to form, because the excited atoms will eventually decay, so that many of the unexcited atoms will become excited. However, suppose the parameters are in the bistable region and the system starts in the ``bright'' state. Then even if one atom decays, it is still effectively on resonance due to its neighbors, and the atom is quickly re-excited instead of forming a domain wall. Thus, in the bistable region, domain walls form at a slower rate than in the antiferromagnetic region, which explains why the correlation decays slower in the bistable region.

The increase of fluctuations in the bistable region is also captured in the excitation susceptibility, which is defined in terms of $E$ as follows, 
\begin{equation}
\chi=N(\langle E^2\rangle-\langle E\rangle^2). %
\end{equation}
The susceptibility measures the correlation averaged over all the spatial degrees of freedom. Figure \ref{fig:sus} (a) plots the susceptibility as a function of $\Omega$ for various system sizes for $V=\Omega=20$. 
The susceptibility is very small for small $\Omega$ and increases as $\Omega$ increases and as the system approaches the bistable region. The peak of the susceptibility is reached at the lower bound of the bistable region predicted by the mean-field approximation. The susceptibility then decreases to a finite value as the $\Omega$ increases. Interestingly, in the vicinity of the bistable region, the susceptibility increases as the system size increases. Outside the bistable region, the susceptibility converges to the same value for different system sizes. The scaling of the susceptibility with the system size is shown in Figure \ref{fig:sus} (b), where we fit the value of the susceptibility at various system sizes with a linear function with one fitting parameter. The result indicates a linear scaling of the susceptibility peak as a function of system sizes.  Interestingly, the scaling exponent 1 is the same as the universal critical scaling exponent for first-order phase transitions in equilibrium \cite{fisher1982scaling}. In general, the scaling exponent for first-order phase transitions equals the dimensionality\cite{fisher1982scaling}. The similarity between the bistable region and the first-order transition implies that the susceptibility will increase linearly in one dimensions as the system grows larger and increase quadratically and cubically in two and three dimensions. In Refs.~\cite{Lesanovsky2012,carr2013cooperative}, the analogy between bistability and first-order phase transitions was also drawn. 

\subsection{Quantum correlation}

\begin{figure}[t]
\includegraphics[width=0.4\textwidth]{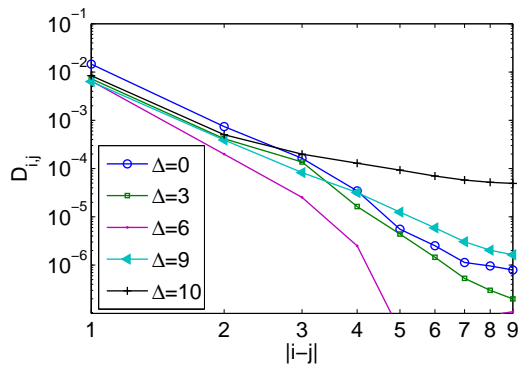}
\caption{Quantum discord between two atoms as a function of distance $|i-j|$ for different $\Delta$ and $V=10$, $\Omega=4.5$. 
}
\label{fig:discord}
\end{figure}
As the result of the dissipation, the steady states of the Rydberg atoms are mixed states which may have classical and quantum correlations. To estimate the quantumness of the correlation among the atoms, we calculate the quantum discord \cite{Zurek2001,Vedral2001} and the entanglement. The calculations are based on the reduced density of two atoms at arbitrary lattice sites. Specifically, the reduced density matrix of atoms at site $i$ and $j$, $\rho_{i,j}$, is given by 

\begin{align}
\rho_{i,j}=&\frac{1}{4}\sum_{k,l=0}^3 R_{k,l}\sigma_i\otimes\sigma_j=\frac{1}{4}\bigg(I_{4\times4}+\sum_{k=1}^3 m_{i,k}\sigma_k\otimes I_{2\times2} \nonumber \\
&+\sum_{k=1}^3 m_{j,k}I_{2\times2}\otimes\sigma_k +\sum_{k,l=1}^3t_{k,l}\sigma_i\otimes\sigma_j\bigg),
\label{eq:rhoij}
\end{align}
where $R_{k,l}=\text{Tr }[\rho(\sigma_k\otimes\sigma_l)]$, $\sigma_0=I_{2\times2}$, $\sigma_i(i=1,2,3)$ are the Pauli matrices, $\vec{m}=\{m_{j,k}\}$ is the three-dimensional Bloch vectors associated with site $j$ and $t_{i,j}$ is the correlation matrix. Our calculation of the quantum discord is based on the formula of the geometric quantum discord derived in \cite{Vedral2010} as,
\begin{equation}
D_{i,j}=\frac{1}{4}\left(\text{Tr }\vec{m_i}^T\vec{m_i}+\text{Tr }t^Tt+k_{max}\right)
\end{equation}
where $k_{max}$ is the largest eigenvalue of matrix $\vec{m_i}^T\vec{m_i}+t^Tt$. The entanglement is measured by the concurrence \cite{Wootters1997,Wootters1998}, as 
\begin{equation}
\mathcal{C}_{i,j}=max\{0,\gamma_1-\gamma_2-\gamma_3-\gamma_4\},
\end{equation}
where the $\gamma_i$ are the eigenvalues in decreasing order of the matrix $R=\sqrt{\rho_{i,j}(\sigma_2\otimes\sigma_2)\rho_{i,j}(\sigma_2\otimes\sigma_2)}$
and the logarithmic negativity \cite{Vidal_negativity} as
\begin{equation}
\mathcal{N}_{i,j}=\mathrm{log}_2||\rho_{i,j}^{\Gamma_i}||_1
\end{equation}
where $\Gamma_i$ is the partial transpose operation over the atom at site $i$ and $||\cdot||_1$ denotes the trace norm.
\begin{figure}[t]
\includegraphics[width=0.4\textwidth]{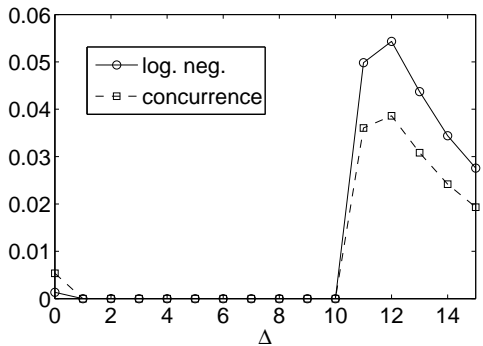}
\caption{Entanglement as measured by logarithmic negativity and concurrence between neighboring atoms as a function of $\Delta$ for different system sizes for $V=10$ and $\Omega=4.5$.} 
\label{fig:entanglement}
\end{figure}

In Figure \ref{fig:discord}, we show the quantum discord as a function of distance for different $\Omega$ with $V=\Delta=10$. The result shows a similar behavior as in the correlation function. For small distances, quantum discord is highest for $\Delta=0$ in the antiferromagnetic region, but the discord decays rapidly as distance increases. The discord decreases as $\Delta$ increases until the system approaches bistability around $\Delta=9$. The decay is the slowest at $\Delta=10$. Different behavior is found in the entanglement as shown in Figure \ref{fig:entanglement}, where we plot the amount of entanglement between nearest neighbors. Entanglement is much stronger in the bistable region than in the antiferromagnetic region.

\section{Conclusion}

In conclusion, we have studied the spatial correlation in various parameter regions for a driven-dissipative 1D lattice of Rydberg atoms. In particular, there is an intriguing slowly decaying correlation in the region of bistability. Such correlations may also occur in other systems with collective bistability like coupled cavities \cite{Ciuti_2012}. It shows the promise of realizing long range correlations in many-body quantum systems with local dissipation. 

We thank Sarang Gopalakrishnan for useful discussions. This work was supported in part by NSF through a grant to ITAMP. 

\bibliography{ryd_corr3}

\end{document}